**Estimation of the Frequency of Occurrence of Italian Phonemes in Text**

J. Arango*, A. DeCaprio*, S. Baik†, L. De Nardis‡, S. Shattuck-Hufnagel† and M.-G. Di Benedetto*†‡

* Radcliffe Institute for Advanced Study at Harvard University, Cambridge, MA, United States
† Massachusetts Institute of Technology (MIT), Cambridge, MA, United States
‡ Sapienza University of Rome, Rome, Italy

*Abstract*

The purpose of this project was to derive a reliable estimate of the frequency of occurrence of the 30 phonemes – plus consonant geminated counterparts- of the Italian language, based on four selected written texts. Since no comparable dataset was found in previous literature, the present analysis may serve as a reference in future studies. Four textual sources were considered: *Come si fa una tesi di laurea: le materie umanistiche* by Umberto Eco, *I promessi sposi* by Alessandro Manzoni, a recent article in *Corriere della Sera* (a popular daily Italian newspaper), and *In altre parole* by Jhumpa Lahiri. The sources were chosen to represent varied genres, subject matter, time periods, and writing styles. Results of the analysis, which also included an analysis of variance, showed that, for all four sources, the frequencies of occurrence reached relatively stable values after about 6,000 phonemes (approx.1,250 words), varying by <0.025%. Estimated frequencies are provided for each single source and as an average across sources.

I.     Introduction

The motivation of this work was to provide an updated dataset of phonemic frequencies in the Italian language. Previously published work on this topic (Zipf and Rogers, 1939; Busa et al., 1962) was completed many decades ago, uses limited sources of text, and relies on particular assumptions about phonemic categories and their classification, particularly in the inclusion of allophones. Specifically, Zipf and Rogers (1939) considers a relatively small corpus of text (5,000 phonemes), and the results may now be outdated. Busa et al. (1962) differentiates between allophonic variants in vowels depending on lexical stress, whereas this project does not consider allophones, in accordance with the phonemic classification in Muljacic (1972). The current study also accounts for the phenomenon of consonant gemination, as will be detailed below. The present dataset may prove useful in further studies of the Italian language for investigations that consider general phonemic tendencies in the language. The dataset may also serve as a benchmark comparison to test whether databases or samples of Italian are representative of the language in general, in terms of phonemic frequencies.

This work was initially developed as a point of comparison for the LaMIT (Lexical access Model for Italian) database of Italian sentences (Di Benedetto et al., 2020) to test whether the text therein was representative of Italian on a phonemic level. The LaMIT Project aims to develop a speech analysis system for lexical access in Italian language based on the model introduced by Ken Stevens (2002), as has been done for English by the Speech Communication Group of the Massachusetts Institute of Technology (MIT), Cambridge, MA, USA. Stevens's model postulates that lexical items are stored in memory as sequences of phonemes defined by hierarchically organized distinctive features. In the model, phonemes are classified based on the distinctive features they possess "such that a change in the value of a feature can potentially generate a new word" (Stevens 2002). Thus, phonemes are determined by their ability to contrast between different word forms. For this reason, this project does not consider allophones that are not contrastive. The classification of phonemes will be detailed in Section II.

Another aim of this study was to investigate the extent to which phonemic frequencies in Italian texts may vary depending on historical period, genre, and whether or not the author is a native speaker. For this reason, the study makes use of four different texts that possess a range of characteristics. The choice of these texts and the method used to transcribe phonemes will be discussed in Section III.

A pertinent question when calculating phonemic frequencies is how large a sample size must be in order to provide reliable and stable results. There are discussions to this end in the past literature on this topic (Busa et al., 1962), but these do not include a quantified justification for the values suggested. A statistical analysis of the rates at which phonemic frequencies converge at the estimated values will be included in Section IV. Graphs showing rates of convergence will be accompanied by a discussion of the minimum size of a corpus necessary to obtain reliable results. Section V will present the results of this study in the estimated frequencies of occurrence for each of the phonemes in the Italian language. A brief discussion, in Section VI, will present further breakdowns of the data and compare the results to values obtained in the past, discussing any notable differences.

**II. Phoneme Classification**

This study was conducted in tandem with the LaMIT (Lexical access Model for Italian) Project (Di Benedetto et al., 2020). The set of phonemes considered in this work, then, is the one determined by the phoneme classification used in LaMIT, which is based on the model introduced by Ken Stevens (2002). In the Stevens (2002) model, it is postulated that words are represented in the mental lexicon in terms of phonemes defined by their distinctive features. Thus, Stevens (2002) provides a classification of the phonemes of the English language according to the distinctive features of each. The LaMIT Project adapts this classification to the Italian language phonemes, using the same set of features. This set of phonemes can be compared to the one presented by Muljacic (1972) as a benchmark. Like the phoneme set suggested by Muljacic (1972), the present one excludes allophones to avoid discrepancies arising from different varieties of pronunciation depending on dialect, accent, and region.

The phoneme set used in this study matches the 30 phonemes identified by Muljacic (1972). Namely, they are the seven vowels: /a/, /i/, /u/, /e/, /ɛ/, /o/, /ɔ/; the twenty-one consonants: /p/, /b/, /f/, /v/, /t/, /d/, /ts/, /dz/, /s/, /z/, /k/, /g/, /d͡ʒ/, /tʃ/, /ʃ/, /m/, /n/, /ɲ/, /l/, /ʎ/, /r/; and the two glides: /j/, and /w/.

In addition to the frequencies of these phonemes, this study also tabulated the frequency of gemination for each of the consonants. The phenomenon of consonant gemination, which is unique to Italian among the Romance languages, offers potential for a wide range of studies (e.g., Esposito and Di Benedetto, 1999; Di Benedetto and De Nardis, 2019a; Di Benedetto and De Nardis, 2019b). Forthcoming research on gemination shows evidence that geminated consonants are the result of the lexical doubling of single consonantal phonemes, and are not indicative of a separate set of geminated consonantal phonemes. Evidence of doubling of single consonants can be seen in an acoustic signal (Di Benedetto, forthcoming), as well as in historical linguistic studies of Latin (Giannini & Marotta, 1989). Therefore, in the tabulation, this study has marked geminates as doubled appearances of the single consonantal phoneme, and also separately provided the frequency with which each consonant is geminated. With this methodology, the frequencies of geminate consonants can easily be derived as well.

There are three phonemes (excluding all vowels) which do not occur in geminated form. These are the consonant /z/ and the glides /j/ and /w/. On the other hand, the consonants /λ/, /ɲ/, /ʃ/, /ts/, and /dz/ always appear in geminated form intervocalically. Since consonants are never geminated in an initial position in standard Italian, except in certain dialects, these phonemes only occur as single phonemes when they begin a word, or when they are followed by a consonant within the word. Note that word initial consonants may, however, become geminated when the phenomenon known as syntactic gemination (in Italian "Raddoppiamento Sintattico") takes place, occurring across words in particular circumstances (Muljacic, 1972). However, syntactic gemination is not accounted for in the present study that is focused on the phonemic level.

**III. Choice of Sources and Method of Transcription**

Past studies (Busa et al. 1962) of phonemic frequencies have made use of a single textual source of Italian. To explore the possibility of differences among authors, time periods, and genres, four sources of Italian text were used to conduct this study: *I promessi sposi* by Alessandro Manzoni (Manzoni, 1827), *Come si fa una tesi di laurea: le materie umanistiche* by Umberto Eco (Eco, 2001), *In Altre Parole* by Jhumpa Lahiri (Lahiri, 2015), and an article from the newspaper *Corriere Della Sera* entitled "Regionali Umbria, I leader e la posta in gioco: cosa può significare per i big l'esito delle elezioni" by Tommaso Labate (Labate, 2019). The sources were chosen to represent various characteristics in order to expand the viability of the sample.

*I promessi sposi* by Alessandro Manzoni, originally published in 1827, was chosen to represent the genre of long-form fiction in a historical time period. *Come si fa una tesi di laurea: le materie umanistiche* by Umberto Eco was chosen to represent the essay genre in a recent time period. *In Altre Parole* by Jhumpa Lahiri was chosen to represent the memoir genre in present day, in addition to being written by a non-native speaker. "Regionali Umbria, I leader e la posta in gioco: cosa può significare per i big l'esito delle elezioni" was chosen to represent present-day journalism. It is clear that considering a single source with a particular characteristic does not allow for generalizations about other texts with that characteristic, but considering a variety of texts does increase the probability that the values estimated in this study are representative of general texts in Italian.

For each of these four sources, the first 6,000 phonemes (equivalent to an average of 1,250 words) were transcribed. A justification for this figure will be provided in the following section. The dataset of transcribed phonemes will be made available online. The phonemes were transcribed in a form of ARPAbet, modified to the Italian language for ease of transcription. The following additions were made to the standard ARPAbet for this project: "GN" to mean /ɲ/, "LH" to mean /λ/, "TS" to mean /ts/, "DZ" to mean /dz/, as well as the doubling of any symbol to represent its geminated form. Table I shows the ARPAbet characters that correspond to the phoneme as notated in the International Phonetic Alphabet (IPA). The text from each source was transcribed manually and the phonemicizations of words were cross-referenced with the *Dizionario di pronuncia Italiana online (DiPI online)*, an online database based on the eponymous dictionary complied by Canepari (Canepari, 2009) and with the Dizionario D'Ortografia e di Pronunzia (Migliorini, Tagliavini, Fiorelli, 1981); When words were unavailable from those sources, the phonemic transcription was cross-checked with the *Collins Italian Dictionary*. Therefore, transcriptions match one of those sources with a single exception: any appearances of an intervocalic "s" were transcribed as the voiceless phoneme /s/ rather than the voiced phoneme /z/. The voiced and voiceless intervocalic "s" are both common and largely interchangeable in spoken Italian, but Canepari provides evidence that the voiceless /s/ can be considered the standard and official intervocalic pronunciation (Canepari 1979), as also supported by the Dizionario D'Ortografia e di Pronunzia (Migliorini, Tagliavini, Fiorelli, 1981). Geminates were tabulated as independent phonemes in order to more easily calculate the rates at which consonants appear in singular and geminated forms.

**IV. Convergence of Estimated Frequency Values**

A relevant concern when calculating estimated phonemic frequency values is how large a dataset must be in order to provide stable and reliable results. Another way of viewing this question is: after how many transcribed phonemes do frequency values vary by acceptably small increments such that the estimated values assure an adequate level of confidence? Busa et al. (1962) state that changes in frequency values are insignificant after 5,000 phonemes. According to Dennis Klatt (Dennis Klatt, personal communication), reliable frequency values can be attained by 1,000 transcribed words, which in this dataset is equivalent to an average of 4,800 phonemes. Naturally, the minimum acceptable size of a dataset depends on how much variation in frequency values is considered acceptable. This study attempts to quantify these values by showing the rates at which phonemic frequencies converge upon stable values.

Figures 1 to 4 show, for each of the four sources considered in this study, how frequency values vary per each increment of 250 phonemes. In other words, for each increment of 250 phonemes, the absolute difference between the frequency value at the beginning and at the end of the increment for each phoneme is plotted on the vertical axis. Figures 1a, 2a, 3a, and 4a show the change in frequency of each individual phoneme in each respective source, including geminated phonemes. Figures 1b, 2b, 3b, and 4b show an average of all the phonemes for each source, with a best-fit power curve overlayed. As indicated by the power curves, rates of phonemic variation decrease exponentially as more phonemes are considered. Rates of change tend to stabilize at relatively low values. For instance, we see that for each of the four sources, by the 3,250-phoneme mark, the change in frequency values for all phonemes remains below ±0.25% for every 250 phonemes added. At the 6,000 phoneme mark, the phoneme with the greatest volatility was the phoneme /s/ in Eco's text, varying by an average of ±0.14% per 250 phonemes. However, most phonemes showed significantly smaller rates of change at the 6,000 mark.

Figure 5 shows the best-fit power curves of the average change in phonemic frequency for each of the four sources. The figure shows that the average volatility of phonemes for each of the sources after 6,000 phonemes were tabulated is 0.025%, nearly an order of magnitude lower than the maximum variation of any individual phoneme. While some difference in the rate of convergence in the four sources appears visible at low values of phonemes tabulated, as more phonemes are considered, the curves are grouped more closely together, suggesting that differences in rates of convergence resulting from particularities of each text fade out when the number of analyzed phonemes is sufficiently high.

## V. Results

The estimated frequency of appearance for each phoneme in each source is visualized in Figure 6 and Table II, in graph- and table-form respectively. As indicated previously, geminated phonemes are counted as doubled appearances of single consonantal phonemes for this study. We see that the vowels /a/, /e/, /i/, and /o/ are the most frequently appearing phonemes, followed by the consonants /n/, /r/, /t/, and /l/. The most commonly occurring phoneme, /a/, appears an average of 11.91% of the time, while the least commonly occurring phoneme, /z/, appears an average of 0.05% of the time.

Table III shows the frequency at which each phoneme occurs in geminated form, on average across all sources. Only the 20 phonemes that can be geminated are shown in this table. We see, for example, that the phoneme /ɲ/, despite only occurring 0.19% of the time (as shown in Figure 6), almost always appeared in geminated form in the texts considered. This is to be expected since /ɲ/ forms part of the group of phonemes that are always geminated intervocalically. In fact, the members of this group, /ʎ/, /ɲ/, /ʃ/, /ts/, and /dz/, proved to be the five most frequently geminated phonemes. For each phoneme that is not a member of this group, the geminated form is less common than the independent form. The most common geminated phoneme is /tt/, occurring with a frequency of 0.91%. The most common phoneme in independent form (of the

20 that can be geminated) is /n/, occurring with a frequency of 7.51%. However, the geminated form of this consonant, /n:/, only occurs with a frequency of 0.18%.

## VI. Discussion

Given these results, the data can be classified to show the frequencies of occurrence of phonemes according to various phonemic categories. Figure 7 and Table IV display the frequencies of vowels, consonants, and glides. The 30 phonemes are broken down into the categories as follows: the seven vowels: /a/, /i/, /u/, /e/, /ɛ/, /o/, /ɔ/; the twenty-one consonants: /p/, /b/, /f/, /v/, /t/, /d/, /ts/, /dz/, /s/, /z/, /k/, /g/, /d͡ʒ/, /tʃ/, /ʃ/, /m/, /n/, /ɲ/, /l/, /ʎ/, /r/; and the two glides: /j/, and /w/. We see that, although the four most common individual phonemes are vowels (/a/, /e/, /i/, /o/, respectively), consonants are more common than vowels as a phonemic class. Glides, on the other hand, appear at an average of 2.83%, by far the lowest of the three phonemic categories.

Figure 8 and Table V present a breakdown of the frequencies of consonantal phonemes according to their manner of articulation. The 21 consonants are classified into types as follows: the six stops: /t/, /d/, /k/, /b/, /g/, /p/; the four fricatives: /s/, /f/, /v/, /ʃ/; the four affricates: /ts/, /dz/, /d͡ʒ/, /tʃ/; the three nasals: /m/, /n/, /ɲ/; and the three liquids: /l/, /r/, /ʎ/. The results show that stops are the most common consonant type, followed by liquids, nasals, fricatives, and affricates, respectively. That stops are the most prevalent type of consonant is perhaps partly a result of the fact that there are more stops (6) than any other consonant type. Nevertheless, the liquids, of which there are only three, make up 12.94% of consonant appearances. This is a result of the relatively high phonemic frequencies of /l/ and /r/.

The current dataset can be compared with past results to verify their soundness. Two past results were considered: Zipf and Rogers (1939), and Busa et al. (1962). Tables VI and VII show the phonemic frequencies obtained in each of the respective studies. As is evident, both studies use sets of phonemes that differ from the one used for this study. For this reason, the results of the studies cannot be compared directly. In particular, Zipf and Rogers (1939) includes the phonemes /kw/ and /gw/, which are absent from the set used for this study, as well as sixteen geminated consonant phonemes (/s:/, /l:/, /t:/, /k:/, /d:/, /n:/, /p:/, /kw:/, /m:/, /r:/, /f:/, /b:/, /d͡ʒ:/, /v:/, /g:/, /tʃ:/). Furthermore, Zipf and Rogers (1939) excludes the phonemes /dz/ and /z/, which are used in this study, and does not distinguish between singular and geminated instances of the four consonants /ts/, /ʎ/, /ʃ/, and /ɲ/. Otherwise, the phoneme set used in Zipf and Rogers (1939) matches the one used in this study.

Busa et al. (1962) includes five allophones used to indicate vowels that receive lexical stress (/á/, /é/, /í/, /ó/, /ú/). Busa et al. (1962) also includes fifteen geminated consonant phonemes (/s:/, /l:/, /t:/, /k:/, /d:/, /n:/, /p:/, /m:/, /r:/, /f:/, /b:/, /d͡ʒ:/, /v:/, /g:/, /tʃ:/), and does not distinguish between singular and geminated instances of the five consonants: /ts/, /ʎ/, /ʃ/, /ɲ/ and /dz/. Otherwise, the phoneme set used in Busa et al. (1962) matches the one used in this study.

To achieve a meaningful comparison between these past results and the current study, the values obtained in the past studies can be modified to fit this study's phonemic classification. Table VIII shows a comparison between this study's results and an adjusted version of the results from Zipf & Rogers (1939) and Busa et al. (1962). The following steps were taken to adjust the earlier values to fit the current phoneme classification. For Zipf and Rogers (1939), (1) the frequency of each geminate phoneme was doubled and added to the frequency of the independent phoneme (to account for geminated phonemes being counted as doubled consonants in the present study), (2) the frequencies of /kw/ and /gw/ were counted as instances of /k/ and /w/ and /g/ and /w/ respectively, and (3) the frequencies of /dz/ and /z/ were marked as zero since they were absent from the phoneme set. Finally, a new set of frequency percentages was calculated accounting for these modifications. These values are found in Figure 14.

For Busa et al. (1962), (1) the frequency of each geminate phoneme was doubled and added to the frequency of the independent phoneme (to account for geminated phonemes being counted as doubled consonants in the present study), and (2) the frequencies of vowel allophones were added to the frequencies of the standard vowels. Then, a new set of frequency percentages was calculated accounting for these modifications. These values are also found in Figure 14.

Table IX quantifies the comparison between the current results and past results by providing a Pearson correlation coefficient. The results show a high degree of correlation between the three sets of results, while also indicating that the results of the present study bear a higher degree of similarity to the results of Zipf & Rogers (1939) than to those of Busa et al. (1962).

## VII. Conclusion

The purpose of this study was to obtain updated values for the frequency of occurrence of the phonemes of the Italian language in written texts. This was done by tabulating the number of appearances of each phoneme in four distinct texts. The set of phonemes was determined largely in accordance with Muljacic (1972), taking into account contemporary research on the phenomenon of gemination. An analysis of the rates at which phonemic frequencies stabilize at their estimated values was conducted for each source, to determine a suitable size for a corpus of phonemes. It was found that all four sources converged upon their frequency values at relatively similar rates, and that the average variation per 250 phoneme interval, after 6,000 phonemes were tabulated, was below 0.025%. This was deemed an acceptable degree of volatility considering the estimated frequency values obtained. The frequency results showed that the vowels /a/, /e/, /i/, and

/o/ were the most frequently appearing phonemes, followed by the consonants /n/, /t/, /r/, and /l/. An analysis of gemination showed that the phonemes that are always geminated intervocalically were (naturally) the most frequently geminated. A further classification of the data by manner features revealed that consonants were more common than vowels or glides, and that among consonants, stops were the most prevalent type. Finally, a comparison of the present results with past studies showed a high degree of correlation, indicating that the phonemic frequency values obtained herein are generally reliable. This present dataset can serve to provide benchmark values of the frequency of occurrence of each phoneme of the Italian language in text. Furthermore, it can be used to further categorize the results to obtain conclusions about specific phonemic tendencies. The present dataset may also be used to determine, in future studies, whether samples of Italian text are representative of the phonemic distribution of the language in general.

Table I - List of IPA phonemes used in study. Columns show each IPA phoneme beside its corresponding ARPAbet translation.

| IPA Phoneme | ARPAbet Phoneme | IPA Phoneme | ARPAbet Phoneme |
|---|---|---|---|
| /a/ | AA | /k/ | K |
| /e/ | EY | /kk/ | KK |
| /ɛ/ | EH | /g/ | G |
| /o/ | OW | /gg/ | GG |
| /ɔ/ | AO | /t/ | T |
| /i/ | IY | /tt/ | TT |
| /u/ | UW | /d/ | D |
| /l/ | L | /dd/ | DD |
| /ll/ | LL | /f/ | F |
| /ʎ/ | LH | /ff/ | FF |
| /ʎʎ/ | LHLH | /v/ | V |
| /r/ | R | /vv/ | VV |
| /rr/ | RR | /s/ | S |
| /j/ | Y | /ss/ | SS |
| /w/ | W | /z/ | Z |
| /n/ | N | /ʃ/ | SH |
| /nn/ | NN | /ʃʃ/ | SHSH |
| /m/ | M | /tʃ/ | CH |
| /mm/ | MM | /tʃtʃ/ | CHCH |
| /ɲ/ | GN | /dʒ/ | JH |
| /ɲɲ/ | GNGN | /dʒdʒ/ | JHJH |
| /p/ | P | /ts/ | TS |
| /pp/ | PP | /tsts/ | TSTS |
| /b/ | B | /dz/ | DZ |
| /bb/ | BB | /dzdz/ | DZDZ |

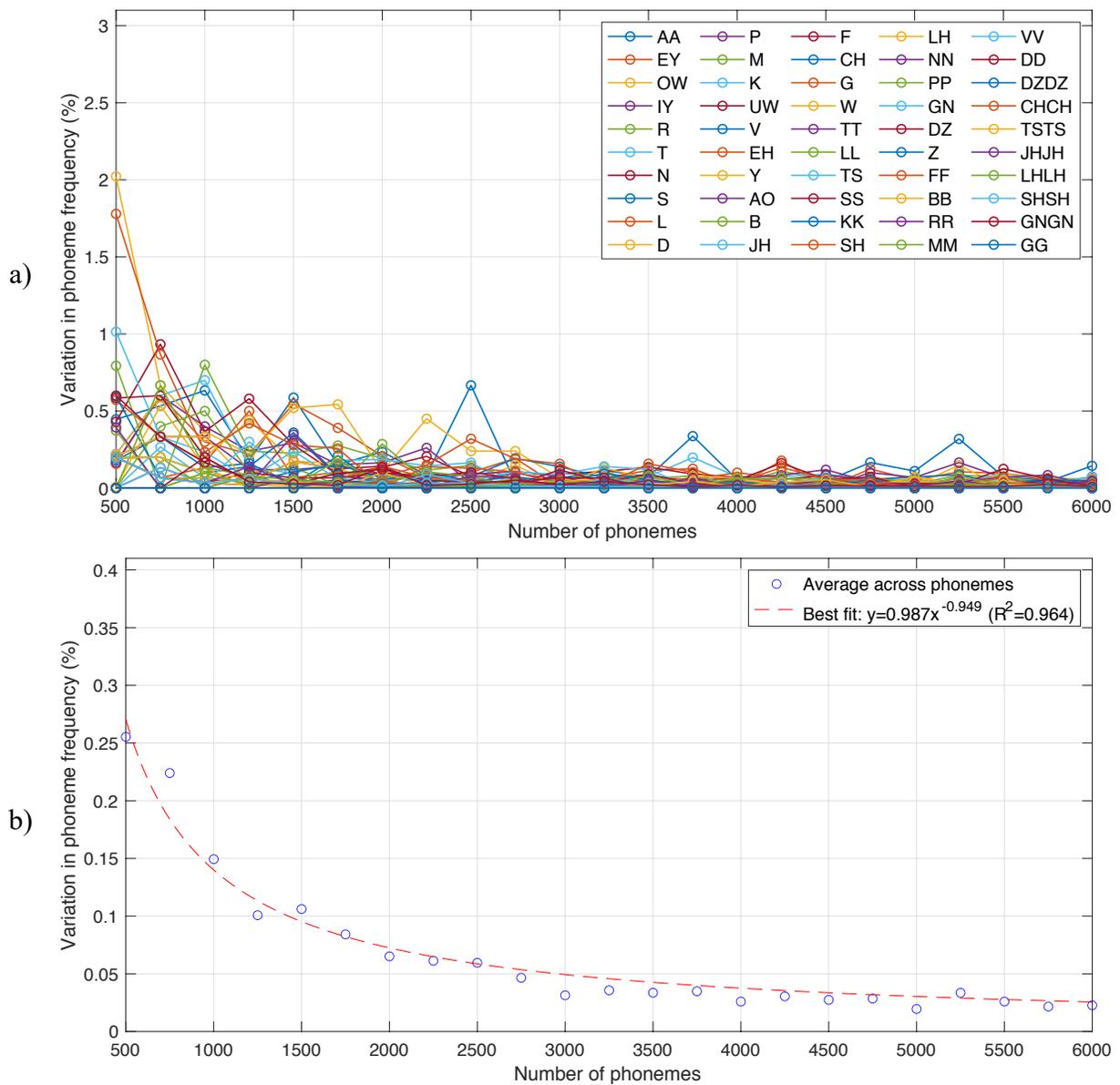

Figure 1 – Change in frequency value for each phoneme per every 250 phonemes interval in the first 6,000 phonemes of Eco's Come si fa una tesi di laurea: le materie umanistiche. Vertical axis charts the change in percentage of a phoneme's frequency value in an interval of 250 phonemes. Value is derived as absolute value of the difference between frequency value at the beginning and end of each interval. Horizontal axis shows number of phonemes considered chronologically in Eco's text. Figure 1a) shows the change in frequency of each individual phonemes in Eco's text, including geminates. Figure 1b) shows the average change in frequency of all phonemes, along with a best-fit power curve with equation $y = 0.987x^{-0.949}$ and $R^2$ value of 0.964. Rates of change decrease for all phonemes as the number of total phonemes considered increases, tending to converge at a relatively low value.

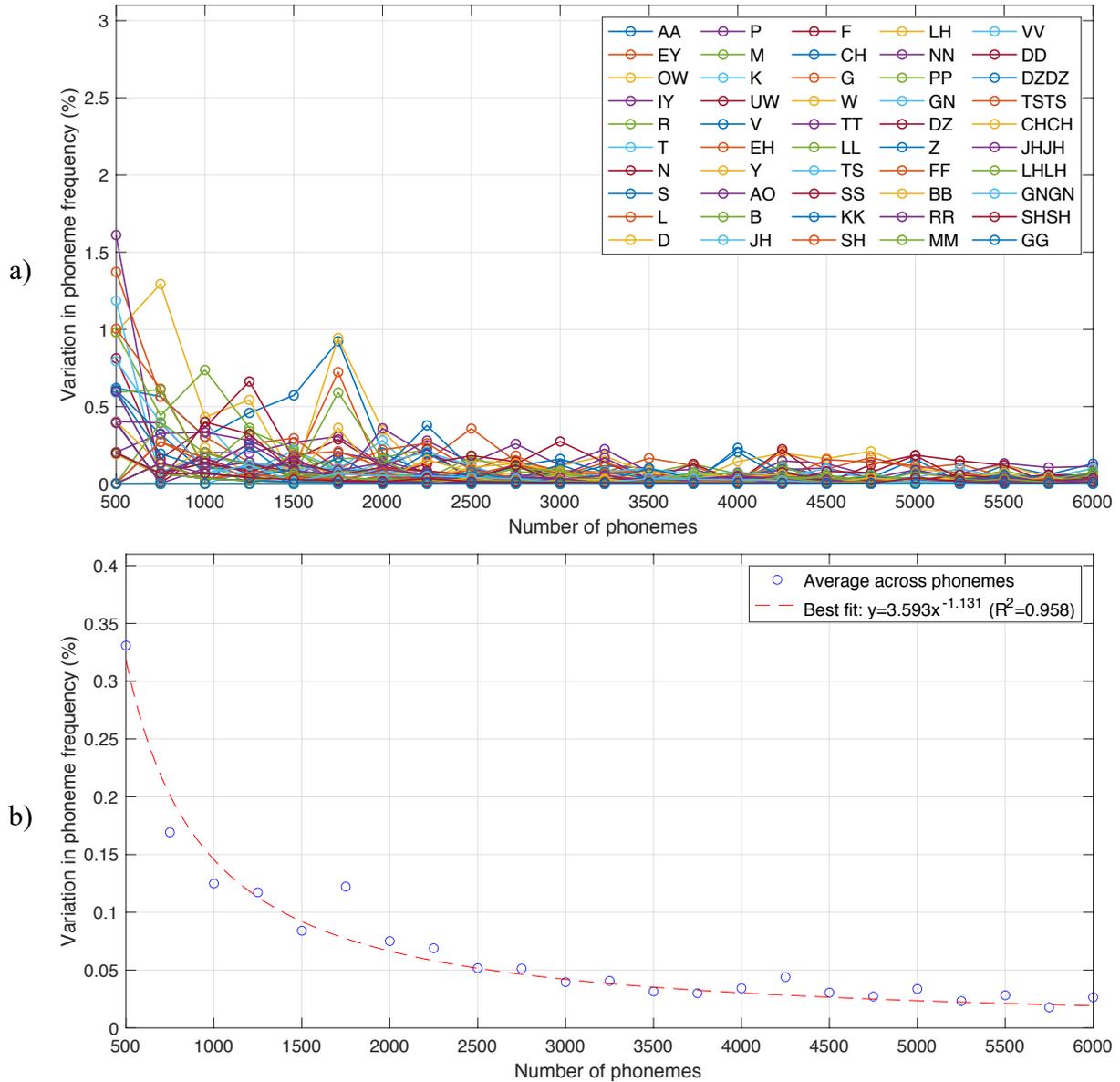

Figure 2 – Change in frequency value for each phoneme per every 250-phoneme interval in the first 6,000 phonemes of Manzoni's I promessi sposi. Vertical axis charts the change in percentage of a phoneme's frequency value in an interval of 250 phonemes. Value is derived as absolute value of the difference between phoneme's frequency value at the beginning and end of each interval. Horizontal axis shows number of phonemes considered chronologically in Manzoni's text. Figure 2a) shows the change in frequency of each individual phoneme in Manzoni's text, including geminates. Figure 2b) shows the average change in frequency of all phonemes, along with a best-fit power curve with equation y = $3.593x^{-1.131}$ and $R^2$ value of 0.958. Rates of change decrease for all phonemes as the number of total phonemes considered increases, tending to converge at a relatively low value.

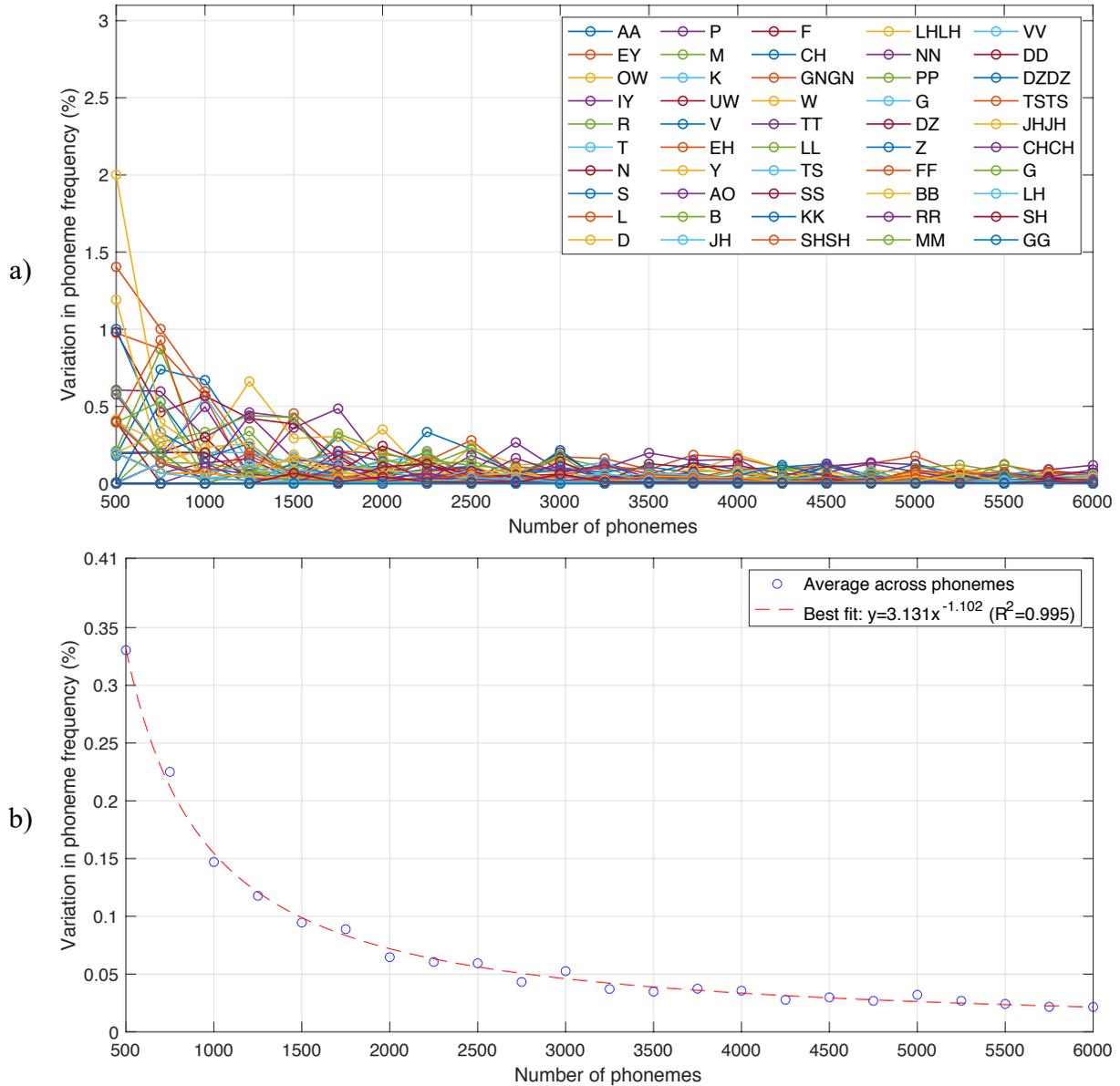

Figure 3 - Change in frequency value for each phoneme per every 250-phoneme interval in the first 6,000 phonemes of Labate's "Regionali Umbria, i leader e la posta in gioco: cosa può significare per i big l'esito delle elezioni". Vertical axis charts the change in percentage of a phoneme's frequency value in an interval of 250 phonemes. Value is derived as absolute value of the difference between phoneme's frequency value at the beginning and end of each interval. Horizontal axis shows number of phonemes considered chronologically in Labate's text. Figure 3a) shows the change in frequency of each individual phoneme in Labate's text, including geminates. Figure 3b) shows the average change in frequency of all phonemes, along with a best-fit power curve with equation $y = 3.131x^{-1.102}$ and $R^2$ value of 0.995. Rates of change decrease for all phonemes as the number of total phonemes considered increases, tending to converge at a relatively low value.

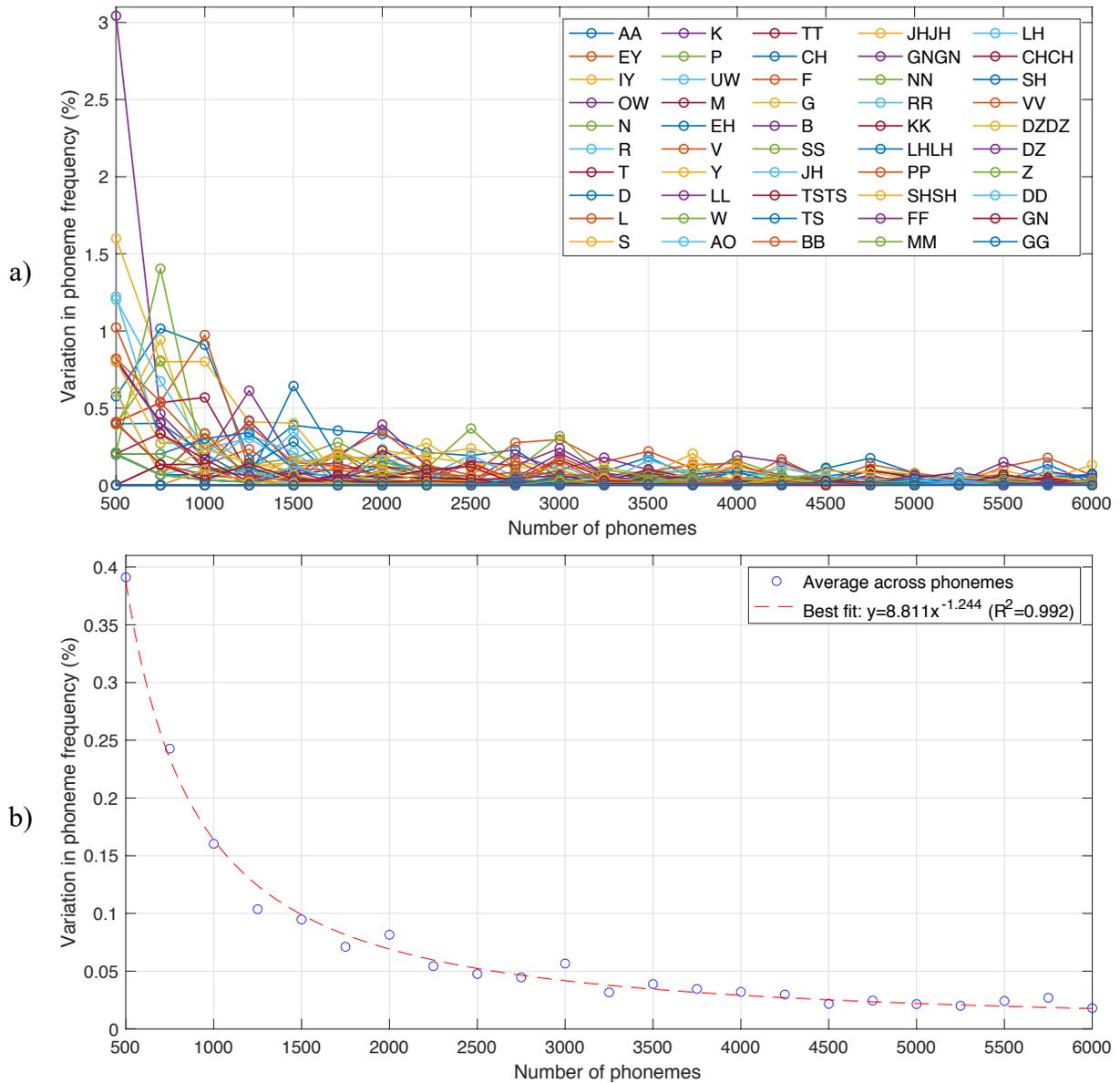

Figure 4 - Change in frequency value for each phoneme per every 250-phoneme interval in the first 6,000 phonemes of Lahiri's *In Altre Parole*. Vertical axis charts the change in percentage of a phoneme's frequency value in an interval of 250 phonemes. Value is derived as absolute value of the difference between phoneme's frequency value at the beginning and end of each interval. Horizontal axis shows number of phonemes considered chronologically in Lahiri's text. Figure 4a) shows the change in frequency of each individual phoneme in Lahiri's text, including geminates. Figure 4b) shows the average change in frequency of all phonemes, along with a best-fit power curve with equation $y = 8.811x^{-1.244}$ and $R^2$ value of 0.992. Rates of change decrease for all phonemes as the number of total phonemes considered increases, tending to converge at a relatively low value.

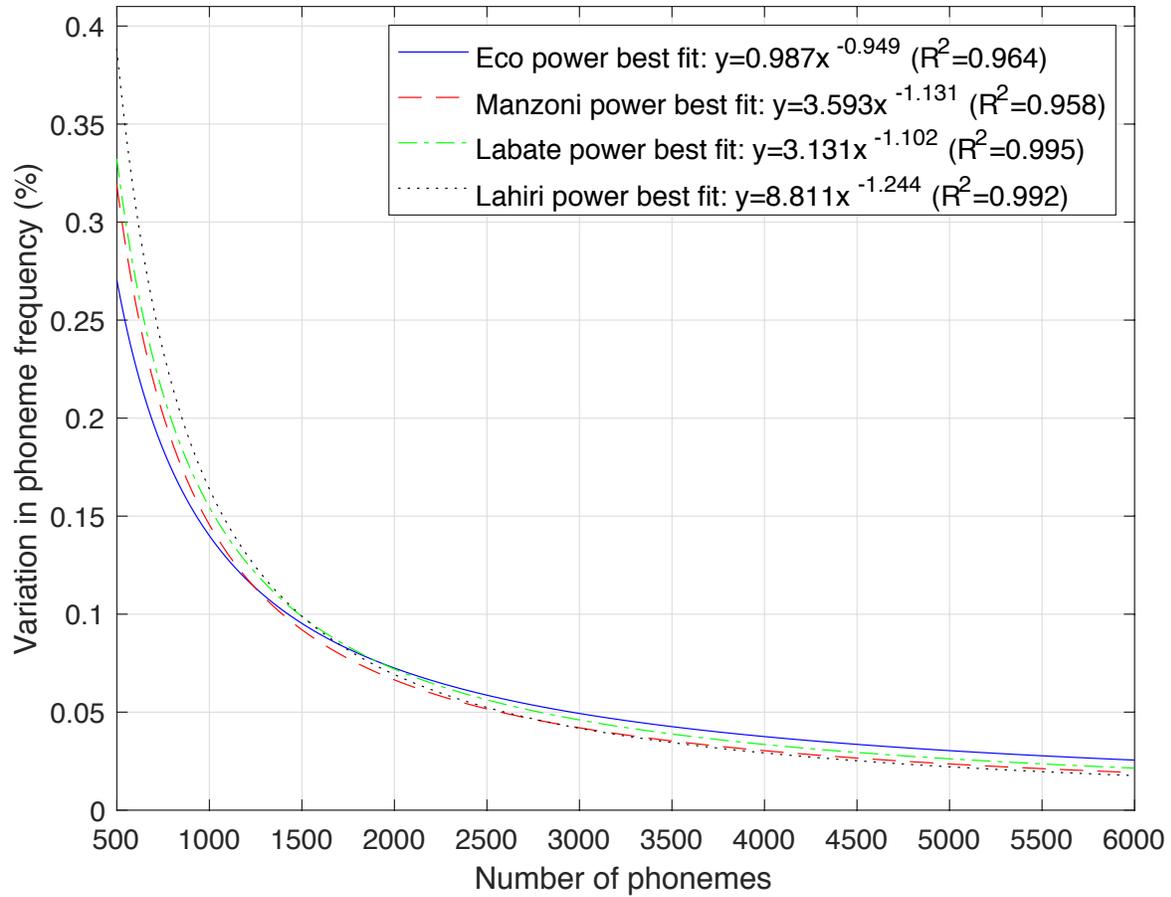

Figure 5 - Power curves of best fit for average rates of change of all phonemes in each 250-phoneme interval for each source. Equation and $R^2$ value for each best fit curve provided in figure legend.

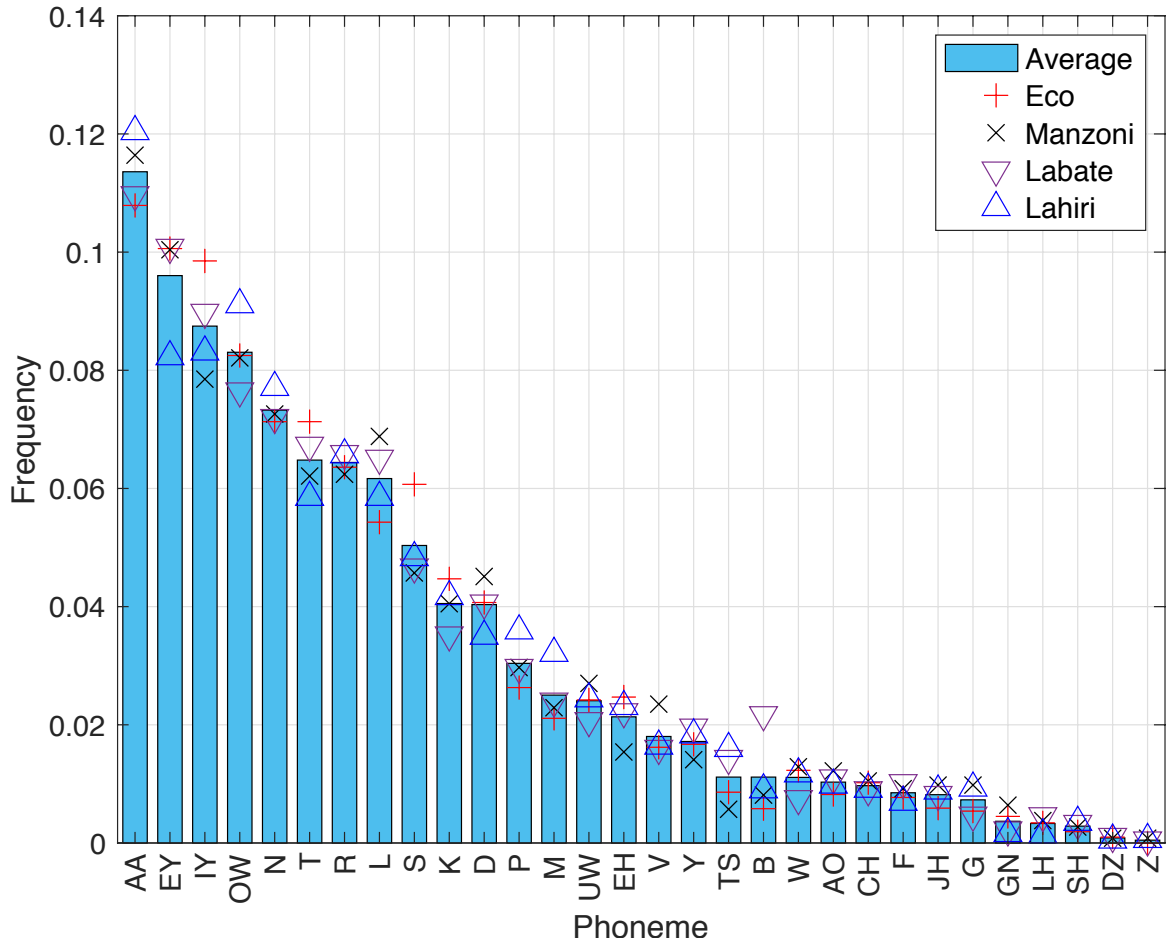

Figure 6 - Values of frequency at which each of 30 phonemes appears in corresponding text after 6,000 phonemes tabulated. Light blue bar represents average frequency of occurrence across all texts. Colored tick marks show frequency value in corresponding source.

| Table II - Phonemic frequencies in table layout. Leftmost column shows ARPAbet translation of all 30 IPA phonemes, followed by columns showing percent occurrence of each phoneme in each respective source. Rightmost column shows average frequency of appearance across four sources. | | | | | |
|---|---|---|---|---|---|
| Phoneme | Percent Appearance (Eco) | Percent Appearance (Manzoni) | Percent Appearance (Labate) | Percent Appearance (Lahiri) | Average Across Sources |
| AA | 10.79% | 11.64% | 10.97% | 12.04% | 11.36% |
| EY | 10.06% | 10.04% | 10.08% | 8.23% | 9.60% |
| IY | 9.85% | 7.85% | 8.98% | 8.31% | 8.74% |
| OW | 8.25% | 8.21% | 7.65% | 9.11% | 8.31% |
| N | 7.13% | 7.26% | 7.20% | 7.71% | 7.32% |
| T | 7.13% | 6.21% | 6.73% | 5.85% | 6.48% |
| R | 6.36% | 6.24% | 6.59% | 6.57% | 6.44% |
| L | 5.43% | 6.88% | 6.51% | 5.85% | 6.17% |
| S | 6.07% | 4.57% | 4.67% | 4.83% | 5.03% |
| K | 4.47% | 4.05% | 3.52% | 4.17% | 4.05% |
| D | 4.07% | 4.51% | 4.06% | 3.50% | 4.03% |
| P | 2.63% | 2.97% | 2.97% | 3.59% | 3.04% |
| M | 2.11% | 2.29% | 2.40% | 3.21% | 2.50% |
| UW | 2.42% | 2.70% | 2.07% | 2.44% | 2.41% |
| EH | 2.47% | 1.54% | 2.22% | 2.31% | 2.13% |
| V | 1.62% | 2.35% | 1.60% | 1.64% | 1.80% |
| Y | 1.67% | 1.41% | 1.96% | 1.83% | 1.72% |
| TS | 0.86% | 0.57% | 1.43% | 1.60% | 1.12% |
| B | 0.58% | 0.81% | 2.17% | 0.90% | 1.11% |
| W | 1.23% | 1.29% | 0.75% | 1.17% | 1.11% |
| AO | 0.82% | 1.22% | 1.10% | 0.98% | 1.03% |
| CH | 1.02% | 1.05% | 0.90% | 0.91% | 0.97% |
| F | 0.77% | 0.92% | 1.02% | 0.69% | 0.85% |
| JH | 0.59% | 0.98% | 0.82% | 0.88% | 0.82% |
| G | 0.54% | 0.98% | 0.47% | 0.93% | 0.73% |
| GN | 0.45% | 0.64% | 0.22% | 0.16% | 0.37% |
| LH | 0.34% | 0.38% | 0.46% | 0.14% | 0.33% |
| SH | 0.19% | 0.27% | 0.33% | 0.35% | 0.29% |
| DZ | 0.10% | 0.08% | 0.11% | 0.05% | 0.08% |
| Z | 0.00% | 0.08% | 0.05% | 0.06% | 0.05% |

| Table III - Leftmost column shows ARPAbet translation of each of 20 IPA phonemes that can be geminated. Respective columns show frequency with which each phoneme appears in independent, and in geminated form, averaged across all four sources. Rightmost column shows frequency with which each phoneme is geminated, derived by dividing percent geminated appearance by total frequency of appearance of the phoneme. | | | |
|---|---|---|---|
| Phoneme | Frequency Independent Appearance | Frequency Geminated Appearance | Frequency of Gemination |
| **GN** | 0.00% | 0.19% | 97.87% |
| **LH** | 0.06% | 0.14% | 69.39% |
| **SH** | 0.06% | 0.12% | 67.45% |
| **TS** | 0.34% | 0.42% | 55.25% |
| **DZ** | 0.03% | 0.03% | 50.02% |
| **JH** | 0.38% | 0.24% | 38.26% |
| **B** | 0.71% | 0.23% | 24.32% |
| **L** | 4.29% | 1.09% | 16.58% |
| **T** | 4.99% | 0.91% | 15.36% |
| **S** | 4.21% | 0.53% | 11.26% |
| **F** | 0.74% | 0.08% | 9.18% |
| **CH** | 0.86% | 0.08% | 8.44% |
| **K** | 3.83% | 0.21% | 5.17% |
| **P** | 2.88% | 0.15% | 5.09% |
| **M** | 2.47% | 0.08% | 3.12% |
| **R** | 6.42% | 0.17% | 2.53% |
| **N** | 7.51% | 0.18% | 2.28% |
| **V** | 1.83% | 0.03% | 1.57% |
| **G** | 0.76% | <0.01% | 0.55% |
| **D** | 4.21% | 0.01% | 0.30% |

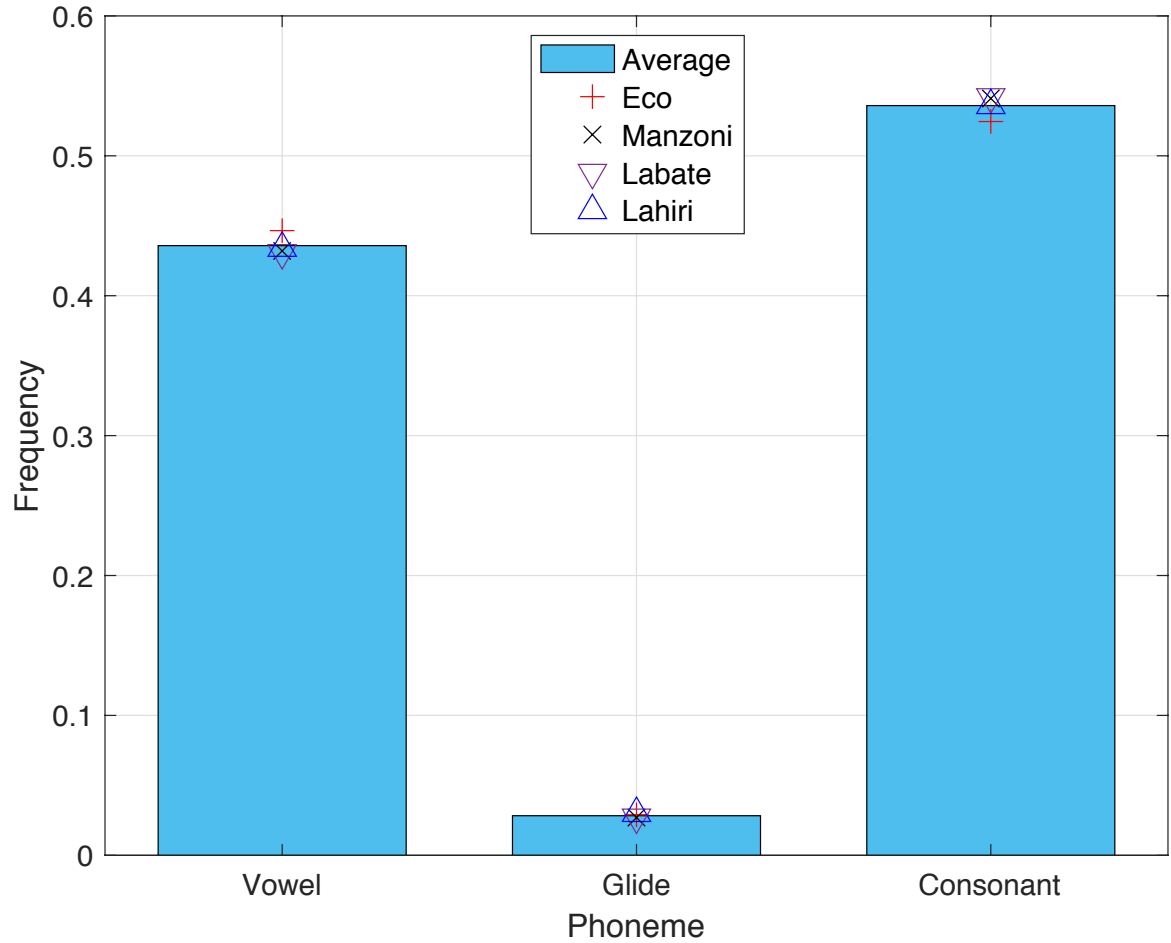

Figure 7 - Frequency of appearance of all 30 phonemes, grouped by phonemic class. Light blue bars represent the average value across all four sources, while colored tick marks represent the values for each individual source. "Vowel" includes: AA, EY, IY, OW, EH, AO. "Glide" includes: Y, W. "Consonant" includes: R, L, LH, N, T, S, D, K, P, M, V, B, CH, F, G, TS, JH, GN, SH, DZ, Z.

| Table IV- Frequency of appearance of phonemes grouped by phonemic class. "Vowel" includes: AA, EY, IY, OW, EH, AO. "Glide" includes: Y, W. "Consonant" includes: R, L, LH, N, T, S, D, K, P, M, V, B, CH, F, G, TS, JH, GN, SH, DZ, Z. | | | | | |
|---|---|---|---|---|---|
| Phoneme Type | Percent Appearance (Eco) | Percent Appearance (Manzoni) | Percent Appearance (Labate) | Percent Appearance (Lahiri) | Average Percent Appearance |
| **Vowel** | 44.65% | 43.20% | 43.07% | 43.40% | 43.58% |
| **Glide** | 2.90% | 2.70% | 2.72% | 3.00% | 2.83% |
| **Consonant** | 52.45% | 54.10% | 54.21% | 53.60% | 53.59% |

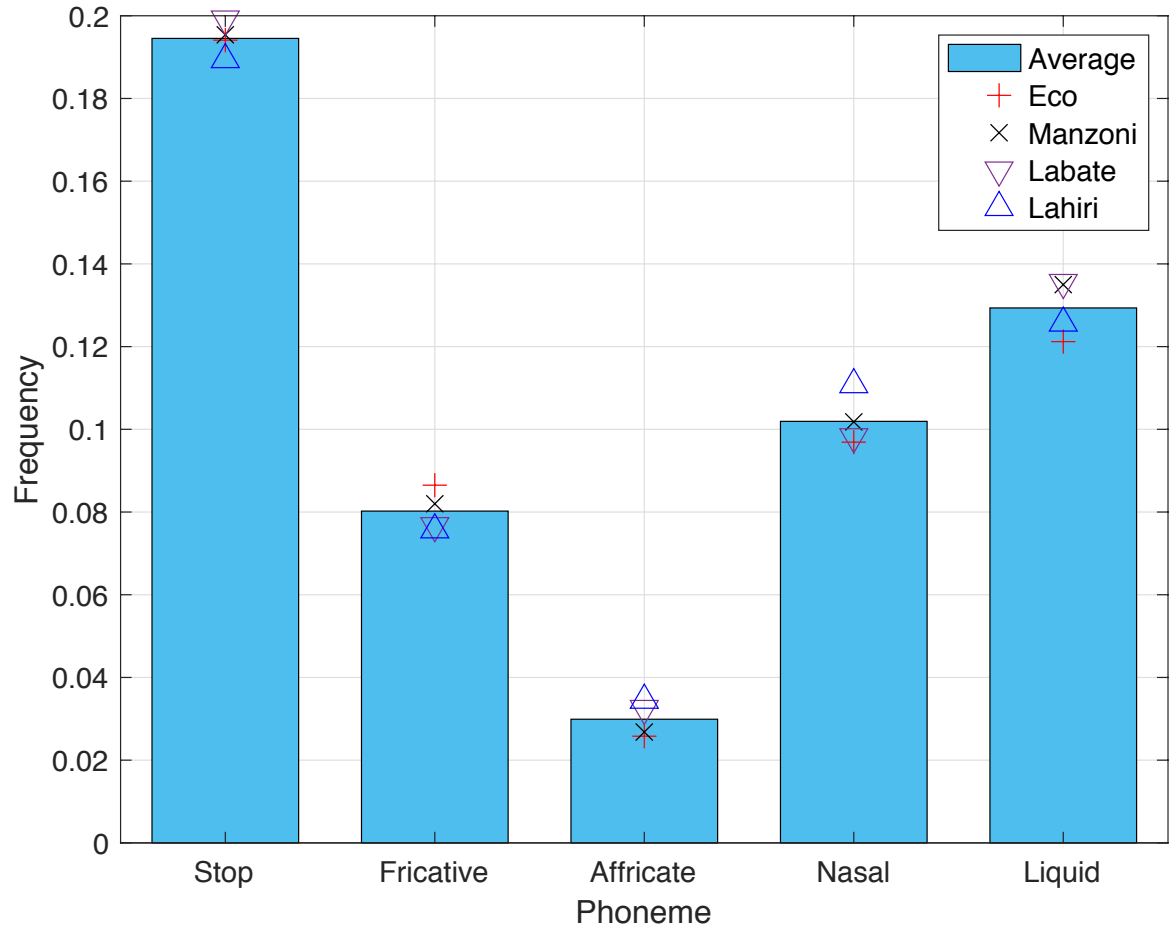

Figure 8 – Frequency of appearance of consonantal phonemes grouped by consonant type (manner feature). Blue bars represent the average frequency value across all four sources, while colored tick marks show the frequency value for each corresponding source. "Stop" includes: T, D, K, P, B, G. "Fricative" includes: S, V, F, SH. "Affricate" includes: CH, TS, JH, DZ. "Nasal" includes: N, M, GN. "Liquid" includes: L, R, LH.

| Table V- Frequency of appearance of consonantal phonemes grouped by consonant type (manner feature). "Stop" includes: T, D, K, P, B, G. "Fricative" includes: S, V, F, SH. "Affricate" includes: CH, TS, JH, DZ. "Nasal" includes: N, M, GN. "Liquid" includes: L, R, LH. | | | | | |
|---|---|---|---|---|---|
| Phoneme Type | Percent Appearance (Eco) | Percent Appearance (Manzoni) | Percent Appearance (Labate) | Percent Appearance (Lahiri) | Average Percent Appearance |
| **Stop** | 19.41% | 19.54% | 19.92% | 18.94% | 19.45% |
| **Fricative** | 8.65% | 8.20% | 7.67% | 7.57% | 8.02% |
| **Affricate** | 2.58% | 2.68% | 3.25% | 3.45% | 2.99% |
| **Nasal** | 9.69% | 10.18% | 9.82% | 11.08% | 10.19% |
| **Liquid** | 12.12% | 13.50% | 13.55% | 12.57% | 12.94% |

| Table VI- Phonemic frequencies obtained by Zipf & Rogers (1939). Values reprinted from Busa et al. (1962). | | | |
|---|---|---|---|
| **Phoneme** | **Frequency** | **Phoneme** | **Frequency** |
| /a/ | 11.46% | /t:/ | 0.70% |
| /e/ | 10.76% | /k:/ | 0.56% |
| /i/ | 9.30% | /d:/ | 0.54% |
| /o/ | 9.20% | /b/ | 0.50% |
| /n/ | 7.60% | /g/ | 0.48% |
| /r/ | 6.30% | /kw/ | 0.46% |
| /t/ | 4.62% | /n:/ | 0.40% |
| /l/ | 4.22% | /p:/ | 0.38% |
| /s/ | 3.74% | /d͡ʒ/ | 0.38% |
| /d/ | 3.48% | /ʎ/ | 0.30% |
| /k/ | 3.32% | /w/ | 0.26% |
| /m/ | 2.52% | /kw:/ | 0.26% |
| /ɛ/ | 2.26% | /m:/ | 0.24% |
| /u/ | 2.18% | /r:/ | 0.22% |
| /p/ | 2.08% | /ʃ/ | 0.22% |
| /j/ | 1.98% | /f:/ | 0.20% |
| /v/ | 1.48% | /ɲ/ | 0.18% |
| /ɔ/ | 1.38% | /gw/ | 0.18% |
| /s:/ | 1.14% | /b:/ | 0.18% |
| /l:/ | 1.10% | /d͡ʒ:/ | 0.14% |
| /f/ | 1.06% | /v:/ | 0.10% |
| /ts/ | 1.00% | /g:/ | 0.10% |
| /tʃ/ | 0.80% | /tʃ:/ | 0.04% |

| Phoneme | Frequency | Phoneme | Frequency |
|---|---|---|---|
| colspan="4" | **Table VII- Phonemic frequencies reported in Busa et al. (1962). Values reprinted from Busa et al. (1962). Vowels with acute accent represent vowels receiving lexical stress.** |||
| /e/ | 8.21% | /tʃ/ | 0.77% |
| /o/ | 8.00% | /t:/ | 0.67% |
| /n/ | 7.27% | /l:/ | 0.65% |
| /r/ | 6.83% | /s:/ | 0.62% |
| /i/ | 6.50% | /b/ | 0.52% |
| /a/ | 6.43% | /ts/ | 0.48% |
| /t/ | 5.67% | /z/ | 0.39% |
| /s/ | 4.42% | /g/ | 0.38% |
| /k/ | 4.10% | /d͡ʒ/ | 0.38% |
| /á/ | 3.96% | /b:/ | 0.25% |
| /d/ | 3.31% | /ʎ/ | 0.21% |
| /l/ | 3.19% | /ʃ/ | 0.20% |
| /m/ | 3.11% | /k:/ | 0.20% |
| /p/ | 2.98% | /ɲ/ | 0.18% |
| /é/ | 2.83% | /r:/ | 0.17% |
| /ó/ | 2.27% | /p:/ | 0.16% |
| /ɛ/ | 2.23% | /m:/ | 0.12% |
| /v/ | 2.13% | /n:/ | 0.12% |
| /j/ | 2.09% | /d͡ʒ:/ | 0.10% |
| /í/ | 1.93% | /tʃ:/ | 0.08% |
| /ɔ/ | 1.38% | /f:/ | 0.05% |
| /ú/ | 1.27% | /v:/ | 0.05% |
| /w/ | 1.18% | /dz/ | 0.02% |
| /u/ | 0.85% | /d:/ | 0.006% |
| /f/ | 0.82% | /g:/ | 0.004% |

| Table VIII- Phonemic frequencies obtained in the current study, compared to adjusted values of past studies to fit the current phoneme classification. | | | |
|---|---|---|---|
| Phoneme | Current | Adjusted Zipf & Rogers (1939) | Adjusted Busa et al. (1962) |
| AA | 11.36% | 10.72% | 10.14% |
| EY | 9.60% | 10.06% | 10.77% |
| IY | 8.74% | 8.70% | 8.22% |
| OW | 8.31% | 8.60% | 10.02% |
| N | 7.32% | 7.85% | 7.33% |
| T | 6.48% | 5.63% | 6.84% |
| R | 6.44% | 6.30% | 6.99% |
| L | 6.17% | 6.00% | 4.38% |
| S | 5.03% | 5.63% | 5.52% |
| K | 4.05% | 5.07% | 4.39% |
| D | 4.03% | 4.26% | 3.24% |
| P | 3.04% | 2.66% | 3.22% |
| M | 2.50% | 2.81% | 3.27% |
| UW | 2.41% | 2.04% | 2.07% |
| EH | 2.13% | 2.11% | 2.18% |
| V | 1.80% | 1.57% | 2.18% |
| Y | 1.72% | 1.85% | 2.04% |
| TS | 1.12% | 0.94% | 0.48% |
| B | 1.11% | 0.80% | 1.00% |
| W | 1.11% | 0.84% | 1.15% |
| AO | 1.03% | 1.29% | 1.35% |
| CH | 0.97% | 0.82% | 0.91% |
| F | 0.85% | 1.37% | 0.90% |
| JH | 0.82% | 0.62% | 0.57% |
| G | 0.73% | 0.80% | 0.38% |
| GN | 0.37% | 0.17% | 0.18% |
| LH | 0.33% | 0.28% | 0.20% |
| SH | 0.29% | 0.21% | 0.20% |
| DZ | 0.08% | 0.00% | 0.02% |
| Z | 0.05% | 0.00% | 0.38% |

| Table IX - Pearson Correlation Coefficients for each of the three datasets in comparison with others. Values approaching 1 indicate high degree of correlation. | | | |
|---|---|---|---|
|  | Current Study | Zipf & Rogers (1939) | Busa et al. (1962) |
| **Current Study** | 1 | 0.993 | 0.981 |
| **Zipf & Rogers (1939)** | 0.993 | 1 | 0.982 |
| **Busa et al. (1962)** | 0.981 | 0.982 | 1 |

## Acknowledgments

This work was supported in part by the Radcliffe Institute for Advanced Study at Harvard University and by Sapienza University of Rome within the research project "Towards Speech Recognition of the Italian Language Based on Detection of Landmarks and Other Acoustic Cues to Features", grant # RP11916B88F1A517.
Stefanie Shattuck-Hufnagel gratefully acknowledges the support of the National Science Foundation, grant # BCS 1827598.

## References

Busa, R., Croatto-Martinolli, C., Croatto, L., Tagliavini, C., Zampolli, A., 1962. Una Ricerca Statistica Sulla Composizione Fonologica Della Lingua Italiana Parlata, Eseguita Con Un Sistema IBM A Schede Perforate. In: XII[th]International Speech and Voice Therapy Conference of the International Association of Logopedics and Phoniatrics, ed. L. Croatto and C. Croatto-Martinolli, Padua.

Canepari, L., 1979. Introduzione Alla Fonetica. Einaudi, p. 197.

Canepari, L., 2009. Il DiPI: Dizionario di Pronuncia Italiana. Zanichelli.

Di Benedetto, M.-G., Choi, J.-Y., Shattuck-Hufnagel, S., De Nardis, L. Budoni, S., Vivaldi, J., Arango, J., DeCaprio, A., Yao, S., 2020. Speech recognition of spoken Italian based on detection of landmarks and other acoustic cues to distinctive features. In: 179th Meeting of the Acoustical Society of America, Chicago, Illinois.

Di Benedetto, M.-G., forthcoming. Lexical and syntactic gemination in Italian. To be submitted.

Di Benedetto, M.-G., De Nardis, L., 2019a. Gemination in Italian: the nasal and liquid case. Submitted to Speech Communication, available at https://arxiv.org/abs/2005.06960.

Di Benedetto, M.-G., De Nardis, L., 2019b. Gemination in Italian: the affricate and fricative case. Submitted to Speech Communication, available at https://arxiv.org/abs/2005.06959.

Eco, U., 2001. Come si fa una tesi di laurea: le materie umanistiche. Bompiani, Florence, Italy, p. 11-14.

Esposito, A., Di Benedetto, M.-G., 1999. Acoustic and Perceptual Study of Gemination in Italian Stops. The Journal of the Acoustical Society of America, ASA, 106(4), 2051-2062. DOI: 10.1121/1.428056.

Giannini, S., Marotta, G., 1989. Fra Grammatica e Pragmatica: La Geminazione Consonantica in Latino. Giardini Editori e Stampatori, Pisa, Italy.

Labate, T., 2019. Regionali Umbria, I leader e la posta in gioco: cosa può significare per i big l'esito delle elezioni. Corriere della Sera, 10/26/19.

Lahiri, J., 2015. In Altre Parole. Guanda, Parma, Italy.

Manzoni, A., 1827. I Promessi sposi. A. Marchese (ed.) 1985, Mondadori, Milan, Italy, p. 5-9.

Migliorini, B., Tagliavini, C., Fiorelli, P., 1981. Dizionario d'ortografia e di pronunzia. RAI ERI, Turin, Italy.

Minnanja, C., Paccagnella, L., 1977. Variazioni dell'entropia linguistica dell'italiano scritto e calcolo di un'entropia fonematica. In: Rendiconti del Seminario Matematico della Università di Padova, 57, 247-265.

Muljacic, Z., 1972. Fonologia della Lingua Italiana. Società editrice il Mulino, Bologna.

Segalina, M., 2012. Dizionari di Pronuncia Italiana Online (ed. L. Canepari). Available at http://www.dipionline.it/dizionario/.

Stevens, K., 2002. Toward a model for lexical access based on acoustic landmarks and distinctive features. The Journal of the Acoustical Society of America, 111(4), 1872-1891.

Zipf, G.K., Rogers, F.M., 1939. Phonemes and Variphones in Four Present-Day Romance Languages and Classical Latin from the Viewpoint of Dynamic Philology. In: Archives néerlandaises de phonétique expérimentale (ANPhE), 15, 111-147.